\documentstyle[prl,aps]{revtex}

\begin{document}
\draft
\twocolumn[\hsize\textwidth\columnwidth\hsize\csname@twocolumnfalse\endcsname
\title{Correlation Functions of Dense Polymers and $c=-2$ Conformal Field Theory}
\author{E.V.~Ivashkevich\cite{Address}}
\address{
Dublin Institute for Advanced Studies,
10 Burlington Road, Dublin 4, Ireland}
\date{January 19, 1997}
\maketitle
\begin{abstract}
The model of dense lattice polymers is studied as an example of non-unitary
Conformal Field Theory (CFT) with $c=-2$. ``Antisymmetric'' correlation 
functions of the model are proved to be given by the generalized Kirchhoff 
theorem. Continuous limit of the model is described by the free complex 
Grassmann field with null vacuum vector. The fundamental property of the 
Grassmann field and its twist field (both having non-positive conformal 
weights) is that they themselves suppress zero mode so that their correlation 
functions become non-trivial. The correlation functions of the fields with
positive conformal weights are non-zero only in the presence of the Dirichlet 
operator that suppresses zero mode and imposes proper boundary conditions.
\end{abstract}
\pacs{05.50.+q, 11.25.H}
]

\makeatletter
\global\@specialpagefalse
\def\@oddhead{\hfill DIAS-STP-98-01}
\makeatother

{\it Introduction.}---In spite of the remarkable success of unitary 
CFT's in predicting the critical properties of 
different lattice spin models \cite{BPZ}, the non-unitary theories, 
although of no less importance for statistical physics, so far were not 
fully understood. It already becomes obvious that some of the 
axioms of unitary CFT have to be sacrificed in this case \cite{G}. 
Still, it remains unclear where one has to modify the foundations and not 
to destroy the whole building of CFT. 

The general idea of the Letter is not to study the non-unitary CFT's on 
their own but, instead, to analyze one particular model of dense polymers 
on the lattice whose continuous limit corresponds to the non-unitary 
$c=-2$ CFT. We believe that at least some of the results obtained on this 
way should be universal and applicable to other non-unitary CFT's.

The model of dense polymers actually has a long history, dating back more
then a century, when Kirchhoff proved a beautiful theorem that the number 
of one-components spanning trees (polymers) on the lattice of $N$ sites is 
given by the principal minors of the $N\times N$ matrix of discrete Laplacian
\cite{H,P}. 
Another fundamental  result was due to Fortuin and Kasteleyn \cite{FK,B} 
who observed that the partition function $Z_N$ of the $q$-component Potts 
model can be represented as a dichromatic polynomial that continuously 
depend on $q$. Although the partition function of the model vanishes in 
the formal limit $q\to 0$ owing to zero mode of the discrete Laplacian, 
its derivative with respect to $q$ does not and gives the partition function 
of one-component spanning trees. 

The purpose of the Letter is to show that:
\begin{itemize}
\item[(i)]The $q\to 0$ limit of the 
Potts model can be carried on in two steps. The first, $\lambda\to 0$,
leads to the model of lattice polymers with arbitrary number of components
$\gamma$; the second, $\kappa\to 0$, to their dense phase. 
Although the partition function of the model again vanishes in the limit, 
some ``antisymmetric'' $2\gamma$-point correlation functions survive. 
\item[(ii)]These correlation functions are given exactly by the minors of
rank $(N-\gamma)$ of the Laplacian matrix. These can be rewritten in terms
of integrals over anti-commuting variables and in continuous limit 
coincide with the correlation functions of the free complex Grassmann field.
\item[(iii)] The vacuum vector of the field theory have to be defined as having
zero norm. The fundamental property of the Grassmann field and its twist 
field (both are primary with non-positive conformal weights) 
is that their operator products define the Dirichlet 
operator that suppresses zero mode and imposes proper boundary 
conditions for the primary fields with positive conformal weights.
This does not change other basic principles of CFT and leads to 
the logical and self-consistent theory. 
\end{itemize}
{\it Dense Phase of Lattice Polymers.}---Let lattice ${\cal L}$ has $N$ 
sites labeled $1,2,...,N$. With each site $i$ we associate a spin variable 
$\sigma_i$ which can take $q$ values, say $1,2,...,q$.

Then the average of any operator ${\cal A}(\sigma)$ in the $q$-component 
Potts model we define as (without normalization factor!)
\begin{equation}
\left\langle{\cal A}(\sigma)\right\rangle=\sum_\sigma {\cal A}(\sigma)
\exp\left\{\beta J \sum_{(ij)}
\delta(\sigma_i,\sigma_j)\right\}.
\label{Z}
\end{equation}
Here the $\sigma$-summation is over all the spins $\sigma_1,...,\sigma_N$;
the second summation is over all edges of the lattice. It has been shown that 
$Z_N$ can be expressed as a dichromatic polynomial \cite{FK,B}. To fix
notations we briefly repeat the derivation of the result.
Set $v=\exp(\beta J)-1$, then the partition function can be rewritten as
\begin{equation}
Z_N=\langle 1 \rangle=\sum_\sigma \prod_{(ij)}\left[1+v\delta(\sigma_i,\sigma_j)\right].
\label{Zprod}
\end{equation}
Let $E$ be the number of edges of the lattice ${\cal L}$. Then the summand in 
Eq.(\ref{Zprod}) is a product of $E$ factors. Each factor is the sum of two 
terms: $1$ and $v\delta(\sigma_i,\sigma_j)$, so the product can be 
expanded as the sum of $2^E$ terms. 

Each of these $2^E$ terms can be associated with a bond-graph on the lattice 
${\cal L}$. To do this, note that the term is the product of $E$ factors, one
for each edge. The factor for edge $(ij)$ is either $1$ or 
$v\delta(\sigma_i,\sigma_j)$: if it is the former, leave the edge empty, 
if the later, place a bond on the edge. Do this for all edges $(ij)$. We then 
have a one-to-one correspondence between bond-graphs on ${\cal L}$ and terms 
in the expansion of the product in Eq.(\ref{Zprod}).

Consider a typical bond-graph ${\cal G}$, containing $N$ sites, $L$ bonds,
$\gamma$ connected components and $\omega$ internal cycles.
These are not independent, but must satisfy Euler's relation
\begin{equation}
L+\gamma=N+\omega.
\end{equation}
Then the corresponding term in the expansion contains a factor $v^L$, 
and the effect of delta functions is that all sites within a 
component must have the same spin $\sigma$. Summing over all independent spins 
and over all bond-graphs ${\cal G}$  that can be drawn on ${\cal L}$ we 
obtain \cite{FK,B}
\begin{equation}
Z_N=\sum_{\cal G} q^\gamma v^L.
\label{Zdichromatic}
\end{equation}
Note that here $q$ need not be an integer. We can allow it to be any real 
number and, in particular, to consider formal limit $q\to 0$. 
Since we are going to deal with not only one- but arbitrary $\gamma$-component
spanning trees, we have to treat the limit in a way different from \cite{FK}. 

At first we consider the limit $\lambda,q,v\to 0$ while 
$\kappa=q/\lambda$ and $x=v/\lambda$ remain finite. As a result we obtain the 
partition function of lattice polymers  
\begin{equation}
\tilde{Z}_N=\lim_{\lambda\to 0}\lambda^{-N}Z_N=
\lim_{\lambda\to 0}\sum_{\cal G} \kappa^{\gamma}
\lambda^{\omega} x^L=\sum_{\cal T} \kappa^{\gamma} x^L.
\label{Zfinal}
\end{equation}
Here the last summation is over all bond-graphs ${\cal T}$ that has no 
internal cycles, i.e. $\omega=0$. Such graphs are usually called spanning 
trees (polymers). 
The number of bonds $L$ of the spanning tree is related to the number of its
components $\gamma$ as $L=N-\gamma$. Hence, the partition function can be 
rewritten as
\begin{equation}
\tilde{Z}_N=\sum_\gamma
\kappa^\gamma \sum_{{\cal T}_\gamma}x^L=
\sum_\gamma {\cal N}_\gamma \kappa^\gamma x^{N-\gamma},
\label{Zmtree}
\end{equation}
where symbol ${\cal T}_\gamma$ denotes the set of different 
$\gamma$-component spanning trees and ${\cal N}_\gamma$ is their total number.
To simplify further notations we take $x\equiv 1$ without loss of generality.

The second limit $\kappa\to 0$ leads to the so-called dense phase of the
polymer model. Since $\gamma\geq 1$ the partition function (\ref{Zmtree}) 
obviously tends to zero in this limit. Nevertheless, the correlation
functions do not necessarily vanish. Indeed, repeating all the steps leading 
to Eq.~(\ref{Zmtree}) one can calculate the  following correlation functions
\begin{mathletters}
\begin{eqnarray}
\lim_{\kappa,\lambda\to 0}\left\langle 1 \right\rangle
&=&0,\\
\lim_{\kappa,\lambda\to 0}\left\langle\delta_{kl}\right\rangle
&=&{\cal N}_{(kl)}={\rm const},\\
\lim_{\kappa,\lambda\to 0}
\left\langle
\left|
\matrix{ 
 \delta_{kl}& \delta_{kq}\cr
 \delta_{pl}& \delta_{pq}\cr
}\right|
\right\rangle
&=&{\cal N}_{(kl)(pq)}-{\cal N}_{(kq)(pl)},\\
&\vdots& \nonumber
\end{eqnarray}
\label{ASCF}
\end{mathletters}
Here $\delta_{kl}=({v}/{q})\delta(\sigma_k,\sigma_l)$; 
${\cal N}_{(kl)}$ is the number of one-component spanning trees with
both the sites $k$ and $l$ belonging to the same component (this number, 
obviously, does not depend on the position of the sites); 
${\cal N}_{(kl)(pq)}$ is the number of two-component spanning trees with 
sites $k,l$ belonging to one component and sites $p,q$ to the other; etc.
The antisymmetric combination of $\delta$'s in each $2\gamma$-point 
correlation function is designed to guard against any contribution of 
spanning trees with the number of components less then $\gamma$ (otherwise
this would be divergent). So, only $\gamma$-component spanning trees 
contribute to the $2\gamma$-point correlation function in the limit 
$\kappa\to 0$. 

The importance of these  correlation functions is justified by the 
following result.

{\it Generalized Kirchhoff Theorem}.---Given a lattice ${\cal L}$ with $N$ 
sites labeled $1,2,...,N$, the $N\times N$ matrix of discrete Laplacian 
$\Delta_{ij}$ has the elements: $\Delta_{ii}=$number of edges incident to 
$i$, $\Delta_{ij}=-$number of edges with end points $i$ and $j$. 
The minor $\Delta^{(k)(l)}$ of rank $(N-1)$ is obtained from the matrix 
$\Delta$ by deleting $k$-th column and $l$-th row; similarly, the  minor 
$\Delta^{(kp)(lq)}$ of rank $(N-2)$ is obtained by deleting columns $k,p$ 
and rows $l,q$; etc. Then 
\begin{mathletters}
\begin{eqnarray}
\det\Delta &=&0,\\
\det\Delta^{(k)(l)}&=&{\cal N}_{(kl)}={\rm const},\\
\det\Delta^{(kp)(lq)}&=&{\cal N}_{(kl)(pq)}-{\cal N}_{(kq)(pl)},\\
&\vdots&\nonumber
\end{eqnarray}
\label{KT}
\end{mathletters}
Here one immediately recognizes the ``antisymmetric'' correlation functions 
(\ref{ASCF}).
The standard proof of the original version of the theorem (first two lines of 
the sequence) can be found in Ref.~\cite{H}. Priezzhev \cite{P} proposed an
alternative proof of the original version in the spirit of the combinatorial 
solution of Ising model. His method is simpler and can also be generalized to 
prove all other lines of the sequence (\ref{KT}).

{\it Free Complex Grassmann Field.}---Using the matrix representation
we can reinterpret the partition function of lattice polymers
as being the partition function of some artificial statistical system. 
To this end we define at each site $i$ of the lattice ${\cal L}$ the pair of 
anti-commuting variables $\theta_i$ and $\theta^{*}_i$ (its complex 
conjugate). Then, using Berezin's definition of the integral over 
anti-commuting variables \cite{ID} we can rewrite the determinant of the 
matrix $\Delta$ as
\begin{eqnarray}
\det \Delta&=&\int d\theta^{*}_1...d\theta_N
\exp\sum_{ij}\theta^{*}_i \Delta^{ij}\theta_j \label{BerezinIntegral}\\  
&=&\int d\theta^{*}_1...d\theta_N
\exp\sum_{ij}(\theta^{*}_i-\theta^{*}_j)(\theta_i-\theta_j).
\nonumber
\end{eqnarray}
In continuous limit this partition function defines field theory with the 
action
\begin{equation}
{\cal S}[\theta]=\frac{1}{4\pi}
\int \partial_{\mu}\theta^{*}\partial^{\mu}\theta~ {\rm d}^2 {\bf r}.
\label{Action}
\end{equation}
The average of any operator ${\cal A}[\theta]$ we define as 
\begin{equation}
\langle{\cal A}[\theta]\rangle = \int [{\rm d}\theta^*{\rm d}\theta]~
{\cal A}[\theta]~ 
\exp-{\cal S}[\theta].
\label{Average}
\end{equation}
Then, although the average of the identity operator is equal to zero 
due to the presence of zero mode, all other correlation functions
of the field $\theta$ are non-trivial and can be normalized so that
\begin{mathletters} 
\begin{eqnarray}
\langle 1 \rangle&=&0,\\
\langle\theta^*_1\theta_2\rangle &=&1,\\
\langle\theta^{*}_1\theta^{*}_2\theta_3\theta_4\rangle&=&
\ln\left(\eta^{12}_{34}\right),\label{CFGF4pt}\\
\langle\theta^{*}_1\theta^{*}_2\theta^{*}_3
\theta_4\theta_5\theta_6\rangle&=&\left|
\matrix{ 
 \ln\left( \eta^{12}_{45}\right)& \ln\left( \eta^{12}_{56}\right)\cr
 \ln\left( \eta^{23}_{45}\right)& \ln\left( \eta^{23}_{56}\right)\cr
}\right|,\\
&\vdots &\nonumber
\end{eqnarray}
\label{CFGF}
\end{mathletters}
The field $\theta$ is scalar and its correlation functions depend only on 
the projectively invariant cross-ratios 
\begin{equation}
\eta^{12}_{34}=\left(\frac{r_{13}r_{24}}{r_{14}r_{23}}\right)^2,~\cdots
\label{CrossRatio}
\end{equation}
Here $\theta_1\equiv\theta({\bf r}_1)$; 
$r_{12}\equiv|{\bf r}_1-{\bf r}_2|$. 
These correlation functions are nothing but asymptotics of the ``antisymmetric'' 
correlation functions (\ref{ASCF}) in the continuous limit. 

The surprising thing is that the Grassmann field itself
suppresses zero mode of the Laplacian operator. In spite of this
unusual property it still can be considered as a primary conformal field with 
the weight $h_\theta=0$.

The stress-energy tensor, 
\begin{equation}
T(z)=~:\!\partial\theta^*\partial\theta\!:~=
\lim_{w\rightarrow z}\left\{
\partial\theta^*(z) \partial\theta(w)+\frac{1}{(z-w)^2}\right\},
\end{equation}
satisfies standard operator product expansion,
\begin{equation}
T(z)T(w)=\frac{-1}{(z-w)^4}+\frac{2T(w)}{(z-w)^2}+\frac{\partial T(w)}{z-w}.
\end{equation}
One can assure himself by direct calculation that the stress-energy tensor is 
indeed the generator of conformal transformations in the sense that
for any correlation function 
$\langle X \rangle=\langle\theta^{*}_1\ldots\theta_{2N}\rangle$
from the sequence (\ref{CFGF}) its transformation law is given by
\begin{equation}
\delta_{\epsilon}\langle X \rangle=
\!\oint_C\! dz~ \epsilon(z)~\langle T(z)X\rangle+
\!\oint_C\! d\bar{z}~ \bar{\epsilon}(\bar{z})~\langle \bar{T}(\bar{z})X\rangle.
\end{equation}
The correlation functions of the field $\theta$ 
satisfy the third-order differential equation coming from the condition of 
degeneration of the operator $(1,3)$ with the weight $h_{1,3}=0$ on the 
third level. This equation actually becomes of the second 
order for the field $\partial\theta$ and, in its turn, coincides
with the condition of degeneration of the operator $(2,1)$ with the weight
$h_{2,1}=1$ on the second level. 

The twist field $\sigma(z,\bar{z})$ can be defined with the use of
the standard operator product expansion
\begin{equation}
\partial\theta(z)\sigma(w,\bar{w})\sim 
\frac{\tau(w,\bar{w})}{\sqrt{z-w}}.
\label{TwistOPE}
\end{equation}
Alternatively, on the lattice it can be defined by means of the construction 
similar to that for the disorder operator in Ising model \cite{S}. 

Conformal properties of the twist field $\sigma$ are similar to those 
of the Grassmann field $\theta$. Namely, its correlation 
functions are non-trivial even in the presence of zero mode.
Its correlation functions can be found from the condition of degeneration of 
the operator $(1,2)$ with the weight $h_{1,2}=-1/8$ on the second level
\cite{S}
\begin{mathletters}
\begin{eqnarray}
\langle\sigma_1\sigma_2\rangle&=&\sqrt{r_{12}},\\
\langle\sigma_1\sigma_2\sigma_3\sigma_4\rangle
&=&\pi\sqrt{r_{12}r_{34}}\sqrt{|\eta(1-\eta)|}\nonumber\\
&\times &\left\{F(\eta)\bar{F}(1-\bar{\eta})+
\bar{F}(\bar{\eta})F(1-{\eta})\right\},
\label{CFTF4pt}
\end{eqnarray}
\label{CFTF}
\end{mathletters}
where $F(\eta)=\!~_2F_{1}({\textstyle \frac{1}{2}},
{\textstyle \frac{1}{2}};1;\eta)$; and $\eta=(z_{13}z_{24})/(z_{12}z_{34})$.
Mixed four-point correlation function of the fields $\theta$ and $\sigma$ 
can also be found using standard techniques of CFT
\begin{equation}
\langle\theta^*_1\theta_2\sigma_3\sigma_4\rangle
=2\sqrt{r_{34}}\left\{H(\eta)+\bar{H}(\bar{\eta})\right\}.
\label{MCF4pt}
\end{equation}
Here $H(\eta)=\ln\left(\sqrt{\eta}+\sqrt{\eta-1}\right)$.

This means that both the Grassmann field $\theta$ and its twist field
$\sigma$ can be considered as primary conformal fields with the weights 
$h_\theta=0$ and $h_\sigma=-1/8$ provided that the vacuum 
state has been defined as having zero norm. These fields are unique 
in having both the property and non-positive conformal weights.

{\it Dirichlet Operator and Green Function.}---There is a simple relation
between the four-point correlation function (\ref{CFGF4pt}) and the Green
function of the Laplacian operator. The most straightforward way to understand
this is follows. Let us consider a conducting plane with a current $I=1$
entering the plane at a point ${\bf r}_1$ and leaving it at a point 
${\bf r}_2$. Then the voltage difference between sites ${\bf r}_3$ and 
${\bf r}_4$ on the plane is given by the four-point function 
$\langle\theta^{*}_1\theta^{*}_2\theta_3\theta_4\rangle$.

The Green function of the Laplacian operator with the Dirichlet boundary 
conditions at the point ${\bf r}_0$ can be defined quite similarly. 
Consider the same conducting plane earthed at the point ${\bf r}_0$.
This means that the voltage at this point is always maintained to be equal to
zero. If a current $I=1$ enters the plane at a point ${\bf r}_1$ 
(and leaves it at the earthed point ${\bf r}_0$) then the voltage at a 
site ${\bf r}_2$ is given by the Green function $G_0({\bf r}_1,{\bf r}_2)$. 

The operator ${\cal D}_0$ that corresponds to the earthed point ${\bf r}_0$ 
can, obviously,  be considered as the product of the field 
$\theta_0$ with its complex conjugate $\theta^*_0$ at the same point. 
We will call it the {\it Dirichlet operator} since it imposes the Dirichlet 
boundary conditions on the Grassmann field. 
With the help of this operator the Green function can be represented as 
\begin{equation}
G_0({\bf r}_1,{\bf r}_2)=\langle{\cal D}_0\theta^{*}_1\theta_2\rangle.
\end{equation}

The Dirichlet operator can formally be defined through the following 
operator products
\begin{equation}
{\cal D}_0=
\lim_{1\rightarrow 0}\left\{\theta^{*}_0\theta_1\right\}=
\lim_{1\rightarrow 0}\left\{\frac{\sigma_0\sigma_1}
{\sqrt{r_{01}}}\right\}.
\end{equation}
Its two-point correlation function can be found from the four-point functions
(\ref{CFGF4pt},\ref{CFTF4pt},\ref{MCF4pt}). 
However, one has to be careful merging different points of the four-point
functions because they diverge logarithmically in the limit.
These divergences have absolutely the same nature as those 
present in the Green function in the thermodynamic or continuous 
limit \cite{ID}. To treat them carefully let us first consider
the correlation functions on the lattice with spacing $a$. 
This scale dictates minimal possible distance between different 
merging points. After arbitrary conformal transformation the lattice is no 
longer uniform and the area of any given fundamental square of the 
lattice acquires an additional factor proportional to the metric on the plane:
${\rm d}s^2=g({\bf r}){\rm d}{\bf r}^2$. Finally, we have
(the factor $\sim a^2$ is absorbed into the metric!) 
\begin{equation}
\langle{\cal D}_1\rangle =1,~~~
\langle{\cal D}_1{\cal D}_2\rangle=\ln\frac{(r_{12})^4}{g_1 g_2},~~~\ldots
\label{CFDO}
\end{equation}
Note, that the Dirichlet operator is scalar and its correlation functions are 
projectively invariant. This is natural since it has been defined 
as the product of scalar fields $\theta$ and $\theta^*$. This also suits its 
interpretation as being the operator that determines boundary conditions.

The operator product of the Dirichlet operator with itself and with
the operators $\theta$ and $\sigma$ cannot be completely determined within 
the CFT. Indeed, merging different points of the correlation functions
(\ref{CFGF}) we can only say that
\begin{mathletters}
\begin{eqnarray}
\lim_{1\rightarrow 0}\left\{
\theta_0{\cal D}_1\right\}&=&k~\theta_0,\\
\lim_{1\rightarrow 0}\left\{
\sigma_0{\cal D}_1\right\}&=&k~\sigma_0,\\
\lim_{1\rightarrow 0}\left\{
{\cal D}_0{\cal D}_1\right\}&=&k~{\cal D}_0,
\end{eqnarray}
\end{mathletters}
where $k$ is some constant. From the point of view of the  
``electrical'' interpretation of the operator given above the most 
natural choice would be $k=1$.

As an example of the field with positive conformal weight
let us consider correlation functions of the local energy 
operator 
\begin{equation}
\varepsilon_0=~:\!\partial_\mu\theta^* \partial^\mu \theta\!:~=
\lim_{1\to 0}\{\partial_\mu\theta^*_0 \partial^\mu \theta_1-
4\pi\delta(r_{01})\}.
\end{equation}
This is primary with conformal weight $h_\varepsilon=1$. 
Its correlation functions can be found from Eqs.~(\ref{CFGF}) and
are all trivial, $\langle \varepsilon_1\ldots\varepsilon_N\rangle=0$,
unless we insert the Dirichlet operator 
\begin{mathletters}
\begin{eqnarray}
\langle {\cal D}_0\varepsilon_1 \varepsilon_2\rangle&=&-\frac{8}{(r_{12})^4},\\
\langle {\cal D}_0\varepsilon_1 \varepsilon_2 \varepsilon_3 \varepsilon_4
\rangle&=&\frac{64}{(r_{12}r_{34})^4}\!+\!\frac{64}{(r_{13}r_{24})^4}\!+\!
\frac{64}{(r_{14}r_{23})^4}.
\end{eqnarray}
\end{mathletters}
This property is common to all primary operators with positive conformal
weights.

Let us summarize the results of the Letter. It has been shown that 
the model of the free complex Grassmann field properly describes the
continuous limit of the lattice model of dense polymers only provided 
its vacuum vector has been defined as having zero norm, 
$\langle 0|0\rangle=0$. 
Nevertheless, it is this vacuum that has to be 
considered when one studies the correlation functions of the primary 
fields with non-positive conformal weights ($\theta$ and $\sigma$). 
These correlation functions, (\ref{CFGF}), imply the mode expansion
\begin{equation}
\theta(z,\bar{z})=\chi_0+2\theta_0\ln|z|-\sum_{n\neq 0}\left(
\frac{\theta_n}{n}z^{-n}+\frac{\bar{\theta}_n}{n}\bar{z}^{-n}\right),
\end{equation}
with the commutation relations
\begin{equation}
\{\chi^*_0,\chi_0\}=\Im,~~~\{\theta^*_n,\theta_m\}=
\{\bar{\theta}^*_n,\bar{\theta}_m\}=n~\delta_{n+m},
\end{equation}
where $\langle\Im\rangle=1$ and $\langle\Im^2\rangle=0$. 
Operator $\Im$ is nothing but the coordinate-independent part of the 
Dirichlet operator. It defines yet another null vector, 
$|\star\rangle=\Im|0\rangle$. This can be normalized so that
$\langle\star|0\rangle=1$. Together these two vectors, $|0\rangle$
and $|\star\rangle$, define physical vacuum state for those primary 
fields that have positive conformal weights.

We conclude that the theory is non-trivial only due to the presence
of two different null vectors that are not orthogonal to each other.
This could be a general feature of other non-unitary CFT's.

I would like to thank V.B.~Priezzhev for many stimulating discussions.

{\it Remark.}---After this Letter was written, V.~Gurarie and
M.~Flohr informed me that the first two lines of the sequence (\ref{CFGF}a,b) 
have already appeared in their paper \cite{GFN} where conformal properties 
of the Quantum Hall State have been analyzed. They also considered 
the product of the fields $\theta$ and $\theta^*$ and noticed that the
resulting operator suppressed zero mode. However, there is an important 
difference between the results of Ref.~\cite{GFN} (see also earlier 
references therein) and this Letter. 
Namely, in this Letter it has been shown that for the CFT to be 
consistent with the lattice model of dense polymers the field $\theta$
have to be considered as a primary field with the weight $h_\theta=0$.
Hence, its product with its complex conjugate (the Dirichlet operator) 
has to be invariant under the projective transformations SL(2,C). 
This is important since it is this operator that determines vacuum state 
for the fields with positive conformal weights.
On the contrary, the authors of Ref.~\cite{GFN} did not consider the field 
$\theta$ as conformal and omitted the metric factors in 
Eq.~(\ref{CFDO}). The ``logarithmic'' operator they obtained in this way
was not scalar but possessed more complicated transformation properties.

\end{document}